\documentclass{llncs}

\usepackage{oldlfont,amssymb,epsf,stmaryrd,epsfig}

\def\lra{\longrightarrow}

\def\H{\mbox{\bf \sf H}}
\def\T{\mbox{\bf \sf T}}

\begin{document}
\thispagestyle{empty}
\title{What do we know \\when we know that a theory is consistent ?}
\author{Gilles Dowek}
\date{}
\institute{\'Ecole polytechnique and INRIA\\
LIX, \'Ecole polytechnique,
91128 Palaiseau Cedex, France. \\
{\tt Gilles.Dowek@polytechnique.fr, http://www.lix.polytechnique.fr/\~{}dowek}}
\maketitle

\begin{abstract}
Given a first-order theory and a proof that it is
consistent, can we design
a proof-search method for this theory that fails in finite time when
it attempts to prove the formula $\bot$? 
\end{abstract}

\section{Searching for proofs in a theory}

\subsection{Who knows that higher-order logic is consistent?}

It is well known that higher-order logic can be presented as a
first-order theory, {\em i.e.} that there exists a first-order theory
\H\ and a function $\Phi$ translating closed formulas of higher-order
logic to closed formulas of the language of \H\ such that the sequent
$\vdash A$ is provable in higher-order logic if and only if the
sequent $\H \vdash \Phi A$ is provable in first-order logic (see, for
instance, \cite{DHKHOL}).  Thus, instead of using a proof-search method
specially designed for higher-order logic, such as higher-order
resolution \cite{Huet72,Huet73}, it is possible to use a first-order
method, such as resolution, to search for proofs in higher-order logic.

However, this reduction is inefficient.  Indeed, if we attempt to
prove the formula $\bot$ with higher-order resolution, the clausal
form of the formula $\bot$ is the empty set of clauses, from these
clauses, we can apply neither the higher-order resolution rule
that requires two clauses to be applied, nor any other rule of
higher-order resolution, that all require at least one clause. Thus,
this attempt to prove the formula $\bot$ fails immediately.
In contrast, the axioms of \H\ give an infinite number of
opportunities to apply 
the resolution rule and thus when searching for a proof of $\H
\vdash \bot$, the search space in infinite.  

Thus, we can say that higher-order resolution ``knows'' that
higher-order logic is consistent, because an attempt to prove the
formula $\bot$ fails in finite time, while first-order resolution does
not.

\subsection{A proof-search method for the theory \H}

There is an link between higher-order logic and higher-order
resolution and another link between higher-order logic and the
first-order theory \H. But can we establish a direct link between
higher-order resolution and the theory \H, without referring to
higher-order logic?

The answer is positive because the translation $\Phi$ can be
inverted: there exists
a function $\Psi$ translating closed formulas of the language of \H\
to closed formulas of higher-order logic such that 
the sequent $\H \vdash B$ is provable in first-order logic if and
only if the sequent $\vdash \Psi B$ is provable in higher-order logic.
Thus, a way to search for a proof of a sequent $\H \vdash B$ is to apply 
higher-order resolution to the sequent $\vdash \Psi B$. Thus, independently
of higher-order logic, higher-order resolution can be seen as special
proof-search method for a the first-order theory \H. 

As $\Psi \bot = \bot$ this first-order proof-search method immediately
fails when attempting to prove the sequent $\H \vdash \bot$. Thus,
this method is much more efficient than applying first-order
resolution to the sequent $\H \vdash \bot$ as it ``knows'' that the
theory \H\ is consistent.

\subsection{A proof-search method for a theory $\T$}

Can we generalize this to other theories than $\H$? Given an
arbitrary first-order theory $\T$ and a proof that $\T$ is
consistent, can we always {\em build in} the theory $\T$, {\em i.e.}
exploit the consistency of $\T$ 
to design a proof-search method that fails in finite time when
required to prove the formula $\bot$?

Of course, as we are not interested in the trivial solution that first
tests if the formula to be proven is $\bot$ and then applies any method
when it is not, we have to restrict to proof-search methods
that do not mention the formula $\bot$.

It is clear that the consistency of the theory $\T$ is a necessary
condition for such a method to exist: if $\T$ is inconsistent, a
complete proof-search method should succeed, and not fail, when
attempting to prove the formula $\bot$.  The main problem is to know
if this hypothesis is sufficient.

\section{Resolution modulo}
\label{resolutionmodulo}

\subsection{Resolution modulo}

{\em Resolution modulo} is a proof-search method for first-order logic
that generalizes higher-order resolution to other theories than the
theory \H.

Some axioms of the theory \H\ are equational axioms.
How to build in equational axioms is well-known: we drop 
equational axioms and we replace unification by equational
unification modulo these axioms (see, for instance,
\cite{Plotkin,PetersonStickel}). Equational unification modulo the
equational axioms of $\H$ is called {\em higher-order unification}. 

From a proof-theoretical point of view, this amounts to define a
congruence on formulas generated by the equational axioms and to
identify congruent formulas in proofs. For instance, if we
identify the terms $2 + 2$ and $4$, we do not need the axiom $2 +
2 = 4$ that is congruent to $4 = 4$, but when we substitute the term
$2$ for the variable $x$ in the term $x + 2$, we obtain the term
$4$. We have called {\em deduction modulo} the system obtained by
identifying congruent formulas in proofs.

But not all axioms can be expressed as equational axioms. For
instance, if the axiom of arithmetic $S(x) = S(y) \Rightarrow x = y$ can
be replaced by the equivalent equational axiom $Pred(S(x)) = x$, the 
axiom $\neg 0 = S(x)$, that has no one-point model, cannot be replaced
by an equational axiom. 

Thus, we have extended deduction modulo by identifying some atomic
formulas with not atomic ones. For instance, identifying
formulas with the congruence generated by the rewrite rules $Null(0)
\lra \top$ and $Null(S(x)) \lra \bot$ is equivalent to having the
axiom $\neg 0 = S(x)$.

When we have such rewrite rules operating directly on formulas, 
equational resolution has to be extended.
Besides the resolution rule, we need to add another rule called {\em
Extended narrowing}. 
For instance, if we identify the formula $P(1)$ with $\neg P(0)$, 
we can refute the set of clauses $\{\neg P(x)\}$, 
but to do so, we have to be able to substitute the term $1$ for the
variable $x$ in the clause $\neg P(x)$, deduce the clause
$P(0)$ and conclude with the resolution rule.
More generally, the {\em Extended narrowing} rule allows to narrow any
atom in a clause with a propositional rewrite rule. The
proposition obtained this way must then be put back in clausal form.
Equational resolution extended with this rule is called 
ENAR --- {\em Extended Narrowing and Resolution} --- or {\em
resolution modulo} for short.

When we orient the axioms of \H\ as rewrite rules and use resolution
modulo, we obtain exactly higher-order resolution.

\subsection{Proving completeness}

Proving the completeness of higher-order resolution, and more
generally of resolution modulo, is not very easy. Indeed higher-order
resolution knows that higher-order logic is consistent, {\em i.e.} it
fails in finite time when attempting to prove the formula
$\bot$. Thus, a finitary argument shows that the completeness of
higher-order resolution implies the consistency of higher-order logic,
and by G\"odel's second incompleteness theorem, the completeness of
higher-order resolution cannot be proved in higher-order logic
itself. This explains that some strong proof-theoretical results
are needed to prove the completeness of higher-order resolution, at
least the consistency of higher-order logic. The completeness proof
given by Andrews and Huet \cite{Andrews71,Huet72,Huet73} uses a result
stronger than consistency: the cut elimination theorem for
higher-order logic.

In the same way, the completeness of resolution modulo rests upon the
fact that deduction modulo the considered congruence has the cut
elimination property. Indeed, when the congruence is defined by
rules rewriting atomic
formulas to non-atomic ones, deduction modulo this congruence may have the cut 
elimination property or not. For instance, deduction modulo the rule
$P \lra Q \wedge R$ has the cut elimination property, but not
deduction modulo the rule $P \lra Q \wedge \neg P$ \cite{DowekWerner}
and resolution modulo this second rule is incomplete. 

Is it possible to weaken this cut elimination hypothesis and require,
for instance only consistency? The answer is negative: the
rule $P \lra Q \wedge \neg P$ is consistent, but resolution modulo
this rule is incomplete. More generally, Hermant \cite{Hermant} has
proved that the completeness of resolution modulo a congruence implies 
cut elimination for deduction modulo this congruence.

\subsection{A resolution strategy} 

At least in the propositional case, resolution modulo can be seen as a
strategy of resolution \cite{frocos}.

For instance, consider the rule $P \lra Q \wedge R$. The {\em Extended
narrowing} rule allows to replace an atom $P$ by $Q \wedge R$
and to put the formula obtained this way back in clausal form. With
this rule, from a clause of the form $C \vee P$ we can derive the
clauses $C \vee Q$ and $C \vee R$ and from a clause of the form $C
\vee \neg P$ we can derive the clause $C \vee \neg Q \vee \neg R$.

We can mimic this rule by adding three clauses 
$\neg \underline{P} \vee Q$,
$\neg \underline{P} \vee R$,
$\underline{P} \vee \neg Q \vee \neg R$ 
and restricting the application of the resolution rules as follows:
(1) we cannot apply the resolution rule using two clauses of the this set 
(2) when we apply the resolution rule using one clause of this set 
the eliminated atom must be the underlined atom.
Notice that this set of clauses is exactly the clausal form of the
formula $\underline{P} \Leftrightarrow (Q \wedge R)$. 
This strategy is in the same spirit as hyper-resolution, but the
details are different. 

If we apply the same method with the formula $\underline{P}
\Leftrightarrow (Q \wedge \neg P)$, we obtain the three clauses 
$\neg \underline{P} \vee Q$,
$\neg \underline{P} \vee \neg P$,
$\underline{P} \vee \neg Q \vee P$
with the same restriction and, like resolution modulo, this strategy
is incomplete: it does not refute the formula $Q$.

The fact that this strategy is complete for one system but not for
the other is a consequence of the fact that deduction modulo the rule
$P \lra Q \wedge R$ has the cut elimination property, but not
deduction modulo the rule $P \lra Q \wedge \neg P$.

Understanding resolution modulo as a resolution strategy seems to be
more difficult when we have quantifiers. Indeed, after narrowing an
atom with a rewrite rule, we have to put the formula back in clausal
form and this involves skolemization. 

\section{From consistency to cut elimination}

We have seen in section \ref{resolutionmodulo} that the theory 
$\T = \{P \Leftrightarrow (Q \wedge \neg P)\}$ is consistent, but that 
resolution modulo the rule $P \lra (Q \wedge \neg P)$ is incomplete. 

Thus, it seems that the consistency hypothesis is not sufficient to
design a complete proof-search method that knows that the theory is
consistent. However the rule $P \lra (Q \wedge \neg P)$ is only one
among the many rewrite systems that allow to express the theory $\T$
in deduction modulo. Indeed, the formula $P \Leftrightarrow (Q \wedge
\neg P)$ is equivalent to $\neg P \wedge \neg Q$ and another solution
is to take the rules $P \lra \bot$ and $Q \lra \bot$. Deduction modulo
this rewrite system has the cut elimination property and hence
resolution modulo this rewrite system is complete.  In other words,
the resolution strategy above with the clauses $\neg \underline{P}$,
$\neg \underline{Q}$ is complete and knows that the theory is
consistent. 

Thus, the goal should not be to prove that if deduction modulo a 
congruence is consistent then it has the cut elimination property,
because this is obviously false, but to prove that a consistent
set of axioms can be transformed into a congruence in such a way that
deduction modulo this congruence has the cut elimination property. 
To stress the link with the project of Knuth and Bendix \cite{KB}, we
call this transformation an {\em orientation} of the axioms. 

A first step in this direction has been made in \cite{stacs} following
an idea of \cite{NegriPlato}. Any consistent theory in propositional
logic can be transformed into a polarized rewrite system such that
deduction modulo this rewrite system has the cut elimination
property. 

To do so, we first put the theory \T\ in clausal form and consider
a model $\nu$ of this theory ({\em i.e.} a line of a truth table).

We pick a clause. In this clause there is either a literal
of the form $P$ such that $\nu(P) = 1$ or a literal of the form $\neg
Q$ such that $\nu(Q) = 0$.

In the first case, we pick all the clauses where $P$ occurs
positively 
$P \vee A_{1}, ..., P \vee A_{n}$ and replace these clauses by the formula
$(\neg A_{1} \vee ... \vee \neg A_{n}) \Rightarrow P$.
In the second, we pick all the clauses where $Q$ occurs
negatively 
$\neg Q \vee B_{1}, ..., \neg Q \vee B_{n}$ and
replace these clauses by the formula
$Q \Rightarrow (B_{1} \wedge ... \wedge B_{n})$.
We repeat this process until there are no clauses left. 
We obtain this way a set of axioms of the form
$A_i \Rightarrow P_i$ and $Q_j \Rightarrow B_j$ such that the atomic
formulas $P_i$'s and $Q_j$'s are disjoint. 

The next step is to transform these formulas into rewrite rules and this
is difficult because they are implications and not equivalences. But,
this is possible if we extend deduction modulo 
allowing some rules to apply only to
positive atoms and others to apply only to
negative atoms. This extension of deduction modulo is called {\em
polarized deduction modulo}.
We get the rules $P_i \lra_{+} A_i$ and $Q_j
\lra_{-} B_j$. Using the fact that the $P_i$'s and the $Q_j$'s are
disjoint, it is not difficult to prove cut elimination for deduction
modulo these rules \cite{stacs}.

So, this result is only a partial success because resolution modulo
is defined for non-polarized rewrite systems and orientation
yields a polarized rewrite system. There may be two ways to bridge the
gap: the first is to extend resolution modulo to polarized rewrite
systems. There is no reason why this should not be possible, but this
requires some work. A more ambitious goal is to produce a
non-polarized rewrite system when orienting the axioms. Indeed, the
axiom $P \Rightarrow A$ can be oriented either as the polarized
rewrite rule $P \lra_{-} A$ or as the non-polarized rule
$P \lra (P \wedge A)$, and similarly the axiom 
$A \Rightarrow P$ can be oriented as the rule 
$P \lra (P \vee A)$. But the difficulty here is to prove that
deduction modulo the rewrite system obtained this way has the cut
elimination property. 

Bridging this gap would solve our initial problem for the
propositional case. Starting from a consistent theory, we would build a
model of this theory, orient it using this model, {\em i.e.}
define a congruence and resolution modulo this congruence would be a
complete proof search method for this theory that knows that the
theory is consistent.

But, this would solve only the propositional case and
for full first-order logic, everything remains to be done. 

We have started this note with a problem in automated deduction:
given a theory \T\ and a proof that it is consistent, can we
design a complete proof-search method for \T\ that knows that \T\ is
consistent? We have seen that this problem boils down to a problem
in proof theory: given a theory \T\ and a proof that it is consistent,
can we orient the theory into a congruence such that deduction modulo
this congruence has the cut elimination property?

We seem to be quite far from a full solution to this problem, although
the solution for the propositional case seems to be quite close.

Some arguments however lead to conjecture a positive answer to this
problem: first the fact that the problem seems almost solved for
propositional logic, then the fact that several theories such as
arithmetic, higher-order logic, and some version of set theory have
been oriented. Finally, we do not have examples of theories that can
be proved to be non orientable (although some counter examples exist
for intuitionistic logic). However, some theories still resist to being
oriented, for instance higher-order logic with extensionality or set
theory with the replacement scheme.

A positive answer to this problem could have some impact on automated 
theorem proving, as in automated theorem proving, like in logic in
general, axioms are often a burden.

\end{document}